\documentclass[runningheads,a4paper]{llncs}

\usepackage{amssymb}
\usepackage{graphicx}

\usepackage[english]{babel}
\usepackage[T1]{fontenc}
\usepackage[babel]{csquotes}
\usepackage{amssymb}
\usepackage{color}
\usepackage{lscape}

\usepackage{hyperref}
\urldef{\mailsa}\path|{dgrissa, sylvie.guillaume, mephu}@isima.fr|
\newcommand{\keywords}[1]{\par\addvspace\baselineskip
\noindent\keywordname\enspace\ignorespaces#1}

\begin{document}

\mainmatter  

\title{Combining Clustering techniques and Formal Concept Analysis to characterize Interestingness Measures}

\author{Dhouha GRISSA$^{1}$\and Sylvie GUILLAUME$^{2}$\and Engelbert MEPHU NGUIFO$^{1}$}
\authorrunning{Dhouha Grissa, Sylvie Guillaume, Engelbert Mephu Nguifo}

\institute{LIMOS, UMR 6158 CNRS, Blaise Pascal$^{1}$ and Auvergne$^{2}$ University,\\
Complexe scientifique des C\'ezeaux, 63177 Aubi\`ere cedex, France\\
\mailsa\\
\url{http://www.isima.fr/limos/}}

\toctitle{Lecture Notes in Computer Science}
\tocauthor{Authors' Instructions}
\maketitle

\begin{abstract}
\emph{Formal Concept Analysis \textit{"FCA"} is a data analysis method which enables to discover hidden knowledge existing in data. A kind of hidden knowledge extracted from data is association rules. Different quality measures were reported in the literature to extract only relevant association rules. Given a dataset, the choice of a good quality measure remains a challenging task for a user. Given a quality measures evaluation matrix according to semantic properties, this paper describes how FCA can highlight quality measures with similar behavior in order to help the user during his choice.
The aim of this article is the discovery of Interestingness Measures \textit{"IM"} clusters, able to validate those found due to the hierarchical and partitioning clustering methods \textit{(\textit{AHC} and \textit{k-means})}. Then, based on the theoretical study of sixty one interestingness measures according to nineteen properties, proposed in a recent study, \textit{FCA} describes several groups of measures.}
\keywords{Formal Concept Analysis, Interestingness Measure, Properties, Clustering Techniques}
\end{abstract}

\section{Introduction}

With the development of storing data techniques in modern societies, the user is concerned with the management of large volumes of database. Thus, arises the problem of extracting knowledge from data (KDD, Knowledge Discovering in Databases) introduced by Piatetsky Shapiro \cite{PS}, where the exponential number of generated rules is an impediment to the user who is facing several interesting and uninteresting rules. Given the problem of finding valid association rules, two objective interestingness measures are used, which are \textit{Support} and \textit{Confidence}. However, these measures are not sufficient to extract only the really interesting knowledge and was challenged in many studies such as \cite{Sese}.

Further to the weakness of the \textit{support-confidence} approach, several interestingness measures \cite{Tan04}, \cite{Lallich04}, \cite{Geng2}, \cite{Vaillant06} have been proposed in the literature to judge the relevance of the extracted rules. However, their large number causes the problem of measures selection. Facing the problem of choosing the right interestingness measure to extract valid association rules, formal studies of interestingness measures were proposed in \cite{Geng2}, \cite{Med07}, \cite{Lallich04}, \cite{Guillaum09} for understanding the behavior of different measures. Those studies consist of the evaluation of the interestingness measures according to semantic properties \cite{PS}, \cite{Tan04}, \cite{LMV03}, \cite{Lallich04} judged important to characterize an interesting measure. This assessment will lead to the construction of a matrix that will be the starting point for an interestingness measures clustering. In this article, we focus on Guillaume and al.'s work \cite{Guillaum09}, \cite{Guillaum10} which is extended to about sixty measures. Indeed, in a first time, they evaluate sixty one objective measures according to nineteen properties and in a second time, they apply an unsupervised classification on those measures, based initially on the most popular clustering methods: the agglomerative hierarchical method \textit{AHC} \cite{WAR63} and the partitioning method \textit{K-means} \cite{Mac67}.
This article is then proposed to validate the clustering methods results. This validation remains a challenging task. Different validity indices were proposed in the literature \cite{Kaijun09} to measure the quality of the generated clusters. However, none of them is a winner and the user is supposed able to find another way proving the clusters.

Based on \textit{FCA}, we check the identified groups of measures according to the methods listed above and to highlight interestingness measures with similar behavior in order to help the user during his choice. Thus, we propose in this paper the discovery of interestingness measures clusters, able to validate those found due to the hierarchical and partitioning clustering methods \textit{(\textit{AHC} and \textit{K-means})}, or to highlight new insights.

The second section recalls the basics of Formal Concept Analysis. The third section presents the interestingness measures and defines $19$ semantic properties. The fourth section combines clustering techniques and \textit{FCA} results to characterize interestingness measures.

\section{Basics of Formal Concept Analysis}

\subsection{Formal Concept Analysis}

Formal Concept Analysis \cite{GW99} is an approach capable of discovering and structuring knowledge. In this subsection, we present some basic definitions \cite{Wille82} needed for this paper.

\paragraph{Definition 1:}

Given a formal context $K = (G, M, R)$ consisting of a binary relation $R$ between a set of objects $G$ and a set of attributes $M$ (i.e. $R \subseteq G \times M$). The relation $gRm \Leftrightarrow (g,m) \in R$, is read: \textit{"the object $g$ has the attribute $m$"}.\\

For a set of objects $O \subseteq G$, we define:
   \begin{center}
       $O^{'} := \left\lbrace m \in M | \forall g \in O, (g, m) \in R \right\rbrace  $;
    \end{center}

and for a set of properties $A \subseteq M$, we define:

    \begin{center}
       $A^{'} := \left\lbrace g \in G | \forall m \in A, (g, m) \in R \right\rbrace .$
    \end{center}

\paragraph{Definition 2:}

The pair $(O, A)$ is called a formal concept of the context $(G, M, R)$ by Wille \cite{Wille82}, with $O \subseteq G$, $A \subseteq M$, $O^{'} = A$ and $A^{'} = O$.  The set $O$ is called the extent of the concept $(O, A)$ and $A$ is called its intent.
Formally, let $L$ be the set of concepts of $(G, M, R)$ and $\leq$, the partial order defined between the concepts by:
$(O1, A1) \leq (O2, A2) \Leftrightarrow A1 \subseteq A2 \Leftrightarrow O2 \subseteq O1$.\\

The pair $(L, \leq )$ is called the lattice of context $(G, M, R)$, which can be graphically represented by the Hasse diagram, which is a downward graph for the understanding of conceptual relationships in data.

\subsection{Association rules}

In the following, we describe association rules in terms of Formal Concept Analysis, formulated by \cite{Pas98}, \cite{Zaki98}.

\paragraph{Definition 3:}

An association rule is an implication of the form $X \rightarrow Y$ consisting of two subsets of attributes $X, Y \subseteq M$, called respectively, the antecedent and the consequent of the rule, where $X \cap Y = \emptyset$. We can say that $X$ implies $Y$ if any object with the attributes in $X$ also has the attributes in $Y$.
The support of the rule $X \rightarrow Y$ is $supp(X \rightarrow Y):= \frac{\vert g(X \cup Y)\vert}{\vert G \vert} $ (where $|G|$ is the cardinality of $G$) and its confidence is $conf(X \rightarrow Y):= \frac{supp(X \cup Y)}{supp(X)}$ .

If the confidence of the rule is equal to $1$ $(conf(X \rightarrow Y) = 1)$, then the rule $r$ is called an exact association rule or an implication, otherwise $r$ is called an approximate association rule.

An association rule is valid if and only if its support and confidence are respectively superior to the minimal thresholds, $minsupp$ and $minconf$. So, for mining association rules \cite{Agrawal93}, there are several algorithms in the literature introduced in \cite{Agrawal94} based on the two most used interestingness measures, the \textit{support} and the \textit{confidence}. Unfortunately, the $Support-Confidence$ approach identifies some weaknesses remaining insufficient for measuring the desired quality.

To address this problem, several measures have been proposed in the literature to eliminate some uninteresting rules and many criteria have been defined to design a good interestingness measure.

\section{Interestingness Measures and semantic properties}

In the following, we present interestingness measures reported in the literature to extract only relevant association rules and recall nineteen semantic properties necessary to evaluate the measures.

\subsection{Interestingness Measures}

To identify interesting association rules and to enable the user to focus on what is interesting for him; about sixty interestingness measures \cite{HH01}, \cite{Tan04}, \cite{Feno07} exist in the literature. All of them are defined with at least one of the following parameters: $p(XY)$, $p(\bar{X}Y)$, $p(X\bar{Y})$ and $p(\bar{X}\bar{Y})$, where $p(XY)= \frac{n_{XY}}{n}$ represents the probability of $XY$, and $\bar{X}$ is the negation of $X$.
However, regarding their high number, the choice of the user becomes very awkward. To help him during his choice of the appropriate measure able to extract the "best rules" from very large databases \cite{Agrawal94}, \cite{BMUT97}, \cite{Bayardo98} and responding to his needs, several studies \cite{Tan04}, \cite{Lallich04}, \cite{LMV04}, \cite{GH06} were reported in the literature.
The theoretical studies consist of the proposition of semantic properties in order to characterize and evaluate the interestingness measures. The experimental studies classify the measures according to their experimental behavior.
The main goal of researchers in the domain is then the user assistance to choose the best interestingness measure, able to respond to his needs. For that, formal properties have been developed \cite{PS}, \cite{LMV03}, \cite{Tan04}, \cite{Geng2}, \cite{BGBG05} in order to evaluate the interestingness measures and to help users understanding their behavior. In the following, we present nineteen semantic properties reported in the literature.

\subsection{Semantic properties of measures}

This subsection lists $19$ of the $21$ desired properties for an interestingness measure $m$, formalized in \cite{Guillaum09}. We ignore two of them: \textit{P1} (\textit{Intelligibility or comprehensibility of the measure}) and \textit{P2} (\textit{Easiness to fix a threshold to the rule}) because we think they are very subjective and depend on the user beliefs.

{\bf Property 3 : Asymmetric measure.} \cite{Tan04}, \cite{Lallich04}

As the antecedent and the consequent of the rule $ X \rightarrow Y $ play different roles, it is desirable for measures assessing differently the rules $X \rightarrow Y$ and $Y \rightarrow X$.

{\bf Property 4 : Asymmetric measure in the sense of the conclusion negation } \cite{Lallich04}, \cite{Tan04}

This property takes into account the symmetry in the sense of the conclusion negation. The measure must be able to distinguish between $ X \rightarrow Y $ and $ X \rightarrow \bar{Y}$.

{\bf Property 5 : Measure assessing in the same way $X \rightarrow Y$ and $\bar{Y} \rightarrow \bar{X}$ in the logical implication case.} \cite{Lallich04}

It is preferable to evaluate in the same way $X \rightarrow Y$ and $\bar{Y} \rightarrow \bar{X}$ in the logical implication case. The evaluation of $\bar{Y} \rightarrow \bar{X}$ reaffirms the implicative relationship of $X$ on $Y$.

{\bf Property 6 : Measure increasing function the number of examples or decreasing function the number of counter-examples.} \cite{PS}, \cite{Lallich04}

It is recognized that less a rule has counter-examples, more it is interesting. Therefore, the assessment of the rule's interest can be measured positively according to the high number of examples $n_{XY}$ of the rule, or to the low number of counter-examples $n_{X\bar{Y}}$.

{\bf Property 7 : Measure increasing function the data size} \cite{Geng2}, \cite{Tan04}

As the rule's interest increases according to the data size $n$, it is interesting to visualize the measure's reaction to the dilatation of the data set. However, if the measure increases with the data size $n$ and provides rule-values close to its maximum value; it would lose its discrimination potential.

{\bf Property 8 : Measure decreasing function the consequent/antecedent size.} \cite{Lallich04}, \cite{PS}

Given $n_{X}$, $n_{XY}$ and $n_{X\bar{Y}}$ fixed values, it is interesting to associate the interest of the rule to the size of $Y$: if $n_{Y}$ increases, the measure should decrease. Therefore, the consequent scarcity begets the interest of the premise existence $X$.

{\bf Property 9 : Fixed value $a$ in the independence case.} \cite{Lallich04}, \cite{PS}

A good interestingness measure should have a fixed value $a$ in the independence case. We say that $X$ and $Y$ are independent when the prior realization of $X$ does not change the appearance probability of $Y$ and also when the prior realization of $Y$ does not change the appearance probability of $X$. In this case, the rules $X \rightarrow Y$ and $Y \rightarrow X$ are not relevant since they provide no information even if the confidence has a high value.

{\bf Property 10 : Fixed value $b$ in the logical implication case.} \cite{Lallich04}

The logical rule is a reference situation marking the absence of counter-examples. In fact, a good quality measure must have a fixed value $b$ in the logical implication case (when $p (Y / X) = 1$ or when $X \subseteq Y$).

{\bf Property 11 : Fixed value $c$ in the equilibrium case.} \cite{Blanchard1}

In the situation of equilibrium, when the number of examples is equal to the number of counter examples, a good interestingness measure should have a fixed value $ c$.

{\bf Property 12 : Identified values in the attraction case between $X$ and $Y$.} \cite{PS}

In the attraction case between the antecedent and the consequent of the rule (i.e when $p(XY) > p(X) p(Y)$), it is desirable that an interestingness measure be able to identify the rule-values of the attraction zone.

The value indicated in $property~12$ corresponds to the fixed value in the independence case. From this remark, we deduce that if $property~9$ is not verified, then \textit{property 12} won't too: if $P9(m) = 0$ then $P12 (m) = 0$. This remark is valid to $property~13$ too.

{\bf Property 13 : Identified values in the repulsion case between $X$ and $Y$.} \cite{PS}

In the repulsion case between the antecedent and the consequent of the rule (i.e when $p(XY) < p(X) p(Y)$), it is desirable that an interestingness measure be able to identify the rule-values of the repulsive zone.

{\bf Property 14 : Tolerance to the first counter-examples.} \cite{Lallich04}, \cite{Vaillant06}

To the appearance of the first counter-examples, some authors \cite{Gras04} suggest a constraint on the shape of the curve of an interestingness measure. Some of them suggest the desire to have a slight decrease in the neighborhood of logical implication, rather than rapid or linear decrease. In fact, a nonlinear measure is sensitive to the appearance of counter-examples and then to the noise [16], which reflects the fact that the user may tolerate or not the appearance of few counter-examples without significant loss of rule interest. Measures are then classified between convex, linear or concave shape.

{\bf Property 15 : Invariance in case of expansion of certain quantities.} \cite{Tan04}

The interestingness measure must be invariant according to the dilation of certain quantities: $1)$ when we multiply $n_{XY}$ and $n_{X\bar{Y}}$ by a positive constant $K_{1}$ and $n_{\bar{X}Y}$ and $n_{\bar{X}\bar{Y}}$ by a positive constant $K_{2}$;

$2)$ when we multiply $n_{X\bar{Y}}$ and $n_{\bar{X}\bar{Y}}$ by a positive constant $K_{1}$ and  $n_{XY}$ and $n_{\bar{X}Y}$ by a positive constant $K_{2}$.

{\bf Property 16 : Desired relationship between $X~ \rightarrow ~Y$ and $\bar{X}~ \rightarrow ~Y$ rules.} \cite{Tan04}

An interestingness measure $m$ must be able to distinguish between $X \rightarrow Y$ and $\bar{X}~ \rightarrow Y$. It must verify the following relation: $m(X \rightarrow Y) = - m(\bar{X} \rightarrow Y)$.

{\bf Property 17: Desired relationship between $X~ \rightarrow ~Y$ and $X~ \rightarrow ~\bar{Y}$ antinomic rules.} \cite{Tan04}

An interestingness measure $m$ must be able to distinguish between $X \rightarrow Y$ and $X \rightarrow \bar{Y}$. It must verify the following relation: $m(X \rightarrow Y) = - m(X \rightarrow \bar{Y})$.

We have the following relationship with \textit{property 4}: if $P_{4}(m)=0$ then $P_{17}(m)=0$.

{\bf Property 18: Desired Relationship between $X~ \rightarrow ~Y$ and $\bar{X}~ \rightarrow ~\bar{Y}$ rules.} \cite{Tan04}

An interestingness measure $m$ must be able to distinguish between $X \rightarrow Y$ and $\bar{X}~ \rightarrow ~\bar{Y}$. It must verify the following relation: $m(X \rightarrow Y) = m(\bar{X} \rightarrow \bar{Y})$.

{\bf Property 19: Antecedent size is fixed or random.} \cite{Lallich04}

The premise size is uncertain when the measure is based on a probabilistic model.

{\bf Property 20: Descriptive or statistical measure.} \cite{Lallich04}

A measure is descriptive when its value is invariant in the case of data dilation, when all the numbers are multiplied by the same factor $k$. Otherwise, it is statistical.

{\bf Property 21: Discriminant measure.} \cite{Lallich04}

A discriminant measure allows us to discern the rules, even when the data set is large.

In the following, we propose a validation of the clustering methods results based on the FCA technique.

\section{Combining Clustering Techniques and FCA}

Based on Guillaume and al.'s work \cite{Guillaum09}, \cite{Guillaum10}, which consists of the evaluation of $61$ interestingness measures according to $19$ properties and the construction of a matrix \textit{(61 objects $\times$ 19 attributes)}, we form a formal context. However, to obtain it, we choose only the preferred properties, those validated by the measures and therefore considering only the number "$1$" in the matrix. Except \textit{property 14} (\textit{Tolerance to the first counter-examples}), on which we applied a complete disjunctive coding. We obtained the three following cases for \textit{P14}: \textit{P14.1} (concave measures), \textit{P14.2} (linear measures) and \textit{P14.3} (convex measures) and we keep only the \textit{P14.1} in our formal context, since it's the value often recommended by expert for quality measure.

\subsection{Clustering results with \textit{CAH} and \textit{K-means}.}

We recall firstly the nine groups of measures described in \cite{Guillaum10} and obtained respectively with \textit{AHC} and \textit{K-means} techniques. There are six measures which belong to different clusters considering the two clustering techniques: \textit{\{Gini, mutual information, fukuda, informational gain, interest, recall\}}. These measures are not considered yet.

The results of the clustering techniques listed below are realized using Matlab software, and based on the Euclidean distance between pairs of measures and on Ward distance for the aggregation step.

\begin{itemize}

\item \textit{C1} : \textit{\{Goodman, Variation support, Pearl \}};
\item \textit{C2} : \textit{\{Implication index, Dependency, Prevalence, Coverage, J-measure, Gray and Orlowska's Dependency\}};
\item \textit{C3} : \textit{\{Sebag, Least contradiction, Descriptive-confirm, Ganascia, Laplace, Confidence, Examples and counter-examples \}} ;
\item \textit{C4} : \textit{\{Discriminant Probabilistic index, Kulczynski, F-measure, Jaccard, Cosine, Support \}};
\item \textit{C5} : \textit{\{Negative Reliability, causal confidence, causal confirmed-confidence, Specificity, Leverage, Putative Causal dependency, VT100, precision, causal confirm\}};
\item \textit{C6} : \textit{\{IIE, IIER, II, likelihood link index, IPEE, IP3E\}};
\item \textit{C7} : \textit{\{Yule's Y, Yule's Q, Mgk, Zhang\}};
\item \textit{C8} : \textit{\{Collective strength, Cohen, Piatetsky, Correlation, Novelty, Odds ratio\}};
\item \textit{C9} : \textit{\{One way support, Two way support, Relative risk, Loevinger, Conviction, Pavillon, Bayes factor, Klosgen\}}.

\end{itemize}

\subsection{Validation process with Formal Concept Analysis.}

Using ConExp platform\footnote{http://conexp.sourceforge.net/}, which is a java-based platform developed to build concept lattices, enabling us to have a good visualization of the concept lattices and interpreting the relationships between different concepts. We apply then the formal context (\textit{61 objects and 19 attributes}) on ConExp to be able to visualize the concepts and the groups of measures. We note that the objects of our context are the interestingness measures through which the class hierarchy is accessed. The attributes are the properties of the interestingness measures.

The whole concept lattice holds $338$ concepts where special groups of measures can be discovered. For example, if we want to know the set of interestingness measures validating  the \textit{P3} property, we simply select the concept to which \textit{P3} is attached and all the asymmetric measures will be highlighted.

Moreover, the lattice reveals different groups of measures sharing the same properties. In the following, we try to validate the clustering methods results (\textit{AHC} and \textit{K-means}) using the lattice.

To validate the clusters, we focus on the visualization of the concept lattice: given a cluster $C_k$, for each element in this cluster, we search for similar objects in the lattice and we select all the nodes to which they are attached. By visualizing the selected nodes, we try to find if there exists one concept able to group all the elements of $C_k$. If we find this central concept, we can argue the cluster obtained by the clustering methods; else many questions arise like should we keep this cluster? or should we split or merge it?

It is always possible to find a formal concept that may contain the elements in the cluster. However those elements may be distant in the sense that they may possess different attributes (or properties). Visualizing the neighborhood of a concept in the lattices highlights close concepts and therefore close objects or attributes.

\subsection{Validation of interestingness measures clusters.}

The verification of the clustering techniques results reveals the presence of:

\begin{itemize}
\item \textbf{\textit{clusters easily validated:}} \textit{C1}, \textit{C4} and \textit{C8};
\item \textbf{\textit{clusters hardly validated:}} \textit{C3}, \textit{C6}, \textit{C7} and \textit{C9};
\item \textbf{\textit{clusters may be questionable:}} \textit{C2} and \textit{C5}.
\end{itemize}

In fact, the first group presents the clusters which are easily validated through the lattice. The visualization of its elements highlights their closeness. The second group is the one to which the clusters hardly validated belong i.e., clusters validated after intense visualization, especially when we remark the shift of additional measures to them. The last group presents the clusters that may be questionable. We find difficulty to discern some clusters because we are not able to find all its elements in the same group and therefore we haven't any interpretation for this case, even after the clustering realized by \textit{AHC} and \textit{K-means}.

\paragraph{\textbf{Clusters easily validated:}}

To check \textit{C1}, we remark that the selection of the concept to which \textit{Variation support} measure is attached reveals its presence in the lattice and the closeness of its measures to each others. The shift of the following measures \textit{\{Odds ratio, collective strength, Cohen, Yule's Y, Yule's Q, novelty, correlation, piatetsky\}} to this group is clear in the lattice and we explain it by the fact that all the highlighted measures verify those properties \textit{\{P5, P18 and P21\}}. The visualization of the neighborhood of the selected concept explains the non-membership of those additional measures to \textit{C1}. In fact, the nearest measures are \textit{\{Odds ratio, collective strength, Cohen\}} which are closer to their own group than to \textit{C1}. Otherwise, we can validate the \textit{C1} cluster if its elements are directly linked to each others.

\textit{C4} cluster is the easiest one to verify according to the lattice, it is sufficient to select the concept which acts as the center of the cluster, the one to which \textit{Discriminant Probabilistic index} is attached, to obtain this cluster. Moreover, the lattice reveals two stable subgroups: \textit{\{Kulczynski, Jaccard\}} and \textit{\{Czekanowski-Dice, Cosine\}}.

The following cluster to check is \textit{C8} and the lattice reveals a very good results. All the measures belonging to this cluster are very close to each others (direct link) and are always found in the same group when we select multiple nodes to which \textit{C8} measures are attached. Moreover, the lattice shows that \textit{Precision} or \textit{Accuracy} is in a direct link with \textit{Odds ratio} and it is explained by the fact that they share \textit{\{P4, P5, P6, P7, P8, P21\}}. Then here, the question arises if we have to merge the second subgroup of \textit{C5}, \textit{\{VT100, Precision\}} to \textit{C8} cluster. However, this don't prevent us to validate the common behavior of all C8 measures. The visualization of the lattice reveals also the neighborhood of \textit{C8} cluster to \textit{C1}.

\paragraph{\textbf{Clusters hardly validated:}}

If we continue with \textit{C3} cluster, we note that when we select the concept to which \textit{Least contradiction} is attached, we find all the elements of \textit{C3}, except \textit{Examples and counter-examples index} that we are not able to show with the highlighted measures. However, when we move up by one level in the lattice, we select the \textit{P11} node which is in a direct link with \textit{Least contradiction}, we obtain our group of measures with the presence of \textit{\{IPEE, IP3E, coverage and prevalence\}}. The two last ones are distant from the cluster we are looking for, which argue their non-membership to \textit{C3}. On the contrary, \textit{\{IPEE and IP3E\}} are very close and we explain it by the fact that they verify \textit{P11} property. To recapitulate, we can validate \textit{C3} cluster after an intense visualization of the lattice.

The \textit{C6} statistical measures behave differently in the lattice and they are divided over three groups: the first one reveals the clustering of \textit{\{IIE, IIER\}}, the second one is \textit{\{II, likelihood link index\}} and the last one is \textit{\{IPEE, IP3E\}}. Then three groups are resulted from the lattice if we eliminate the distant measures from each of them. However, at a certain level of the lattice and when we select the concept validating \textit{\{P4, P5, P6, P14.1\}}, we find the cluster \textit{C6} with the presence of \textit{\{Zhang, Yule's Q and Example and counter-example\}}, a set of measures which verify also the same previous properties but are closer to their groups than to \textit{C6}. Consequently, we can validate \textit{C6} cluster.

The same method of visualization is also applied to \textit{C7} cluster and we note that it is not possible to find all its elements in the same group. In fact, we find the two following groups: \textit{\{Yule's Y, Yule's Q\}} and \textit{\{Zhang, Mgk\}} and we visualize through the lattice that there is no direct link between them. For example, \textit{Figure 1} shows that there is no link between \textit{Mgk} and \textit{Yule's Q}, which explains the fact that we don't find them in the same group. Those groups share an important number of properties \textit{\{P4, P5, P6, P7, P9, P10, P12, P13, P17, P21\}} which prove their similar behavior. But, if we search more and try to visualize the core of the lattice, we find that at a certain level of it, there is one central node able to cluster all the \textit{C7} measures. \textit{Figure 1} illustrate this cluster and explains the relationship between the measures and shows how they are linked to each others. The proximity between \textit{C7} measures is also easily readable in the figure and then we can validate \textit{C7} cluster.

\begin{figure}
\centering
\includegraphics[height=6.1cm, width=9cm]{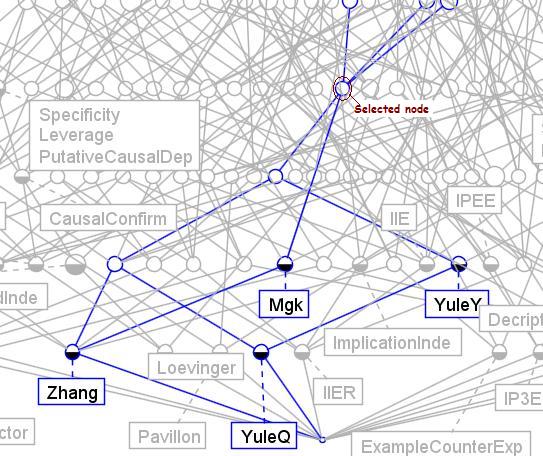}
\caption{C7 cluster}
\label{fig:GL_Yule}
\end{figure}

The last cluster to verify is \textit{C9}, and the concept which acts as center of the group is the one to which "\textit{One Way support}" is attached (\textit{Figure 2}). When we select this node, we remark that all the elements of \textit{C9} are highlighted and we note the presence of some additional measures \textit{\{J-measure, Gray and Orlowska, Dependency, Prevalence and Coverage\}} which share some properties with \textit{C9} measures; some of them are distant but others are near like \textit{\{J-measure, Gray and Orlowska's dependency\}}. The proximity of the latter and their direct link to \textit{Klosgen} don't prove their membership to \textit{C9} because they are close to their own group \textit{C2} too. \textit{Figure 2} shows how the \textit{C9} measures are grouped together, then we can validate it.

\begin{figure}
\centering
\includegraphics[height=6.6cm, width=9cm]{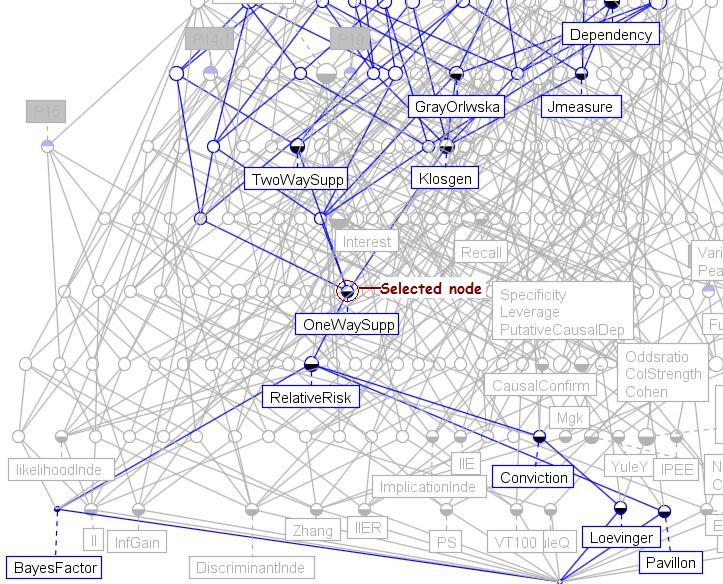}
\caption{C9 cluster}
\label{fig:GL}
\end{figure}


\paragraph{\textbf{Questionable clusters:}}

Concerning \textit{C2} cluster, it is difficult for us to discern it because any of the $6$ measures concepts selected reveal a good results. Only the selection of \textit{Coverage} concept show this group with the presence of all the asymmetric measures, because \textit{Coverage} and \textit{P3} are attached to the same node. Then, the selection of \textit{coverage} node doesn't help us validating \textit{C2}. However, when we select \textit{J-measure} node, we have the following group \textit{\{J-measure, Dependency and Coverage\}}. When we select \textit{Gray and Orlowska's dependency}, we have \textit{\{Gray and Orlowska's dependency, Prevalence and Coverage\}} and finally when we select \textit{Implication index}, we have \textit{\{Implication index, Dependency, Prevalence and Coverage\}}. We remark that there is a link between the \textit{C2} measures according to the three different groups revealed which make this cluster questionable. Consequently, we can't validate \textit{C2} cluster.

The verification of \textit{C5} measures through the concept lattice shows that we are not able to find them in the same group, at least separated between two groups. On the one hand, we find \textit{\{Negative Reliability, causal confidence, causal confirmed-confidence, causal confirm, Specificity, Leverage, Putative causal dependency\}} grouped together; on another hand, we have \textit{\{VT100, Precision\}}. As we visualize in the lattice, property \textit{P3} and \textit{P18} implies \textit{C5} measures separation: the first group doesn't validate \textit{P18}, contrary to the second one. The latter doesn't validate \textit{P3}, contrary to the first group. In addition, we note the presence of external measures to this group, some of them are separated by an important number of concepts and others are very close, like \textit{Loevinger} measure which is directly linked to \textit{Negative Reliability}. Then we can say that \textit{Loevinger} shares some properties with \textit{C5} measures and that it is near to this group of measures even if it belongs to \textit{C9}. Two stable subgroups are revealed from the lattice: \textit{\{Negative Reliability, causal confidence, causal confirmed-confidence\}} and \textit{\{Specificity, Leverage, Putative Causal dependency\}}.
Finally, we remark that all the \textit{C5} measures validate \textit{\{P4, P6, P7, P21\}} and that three of those properties are also shared by \textit{C4}, which explains the closeness of \textit{C4} and \textit{C5} that the hierarchical diagram presented in \cite{Guillaum10} argue too. The question which arises here is if we have to split \textit{C5} cluster into $2$ groups.

\subsection{FCA as a third party}

FCA can also help to determine the cluster of measures that are in different clusters with the two clustering techniques. In our case, six measures \textit{\{recall, informational gain, interest, Gini, mutual information, fukuda\}} are in different clusters. 

Concerning \textit{recall} measure, the selection of the node to which it is attached shows its grouping with both \textit{C4} and \textit{C5} measures. But in term of closeness, it is nearer to \textit{C5}, as it is in a direct link with \textit{specificity}. \textit{Informational gain} is directly linked to \textit{interest} and the latter is grouped with both \textit{C8} and \textit{C9} measures i.e., the lattice is not able to affect those measures to any of those clusters. The same thing with \textit{\{Gini and mutual information\}}, no additional information is revealed concerning them if we remark that graphically, they are near to \textit{C1} like to \textit{C2}. Finally, the selection of \textit{Fukuda} measure reveals that it is close to \textit{C6} measures \textit{\{IIE, IIER and IP3E\}}.

\section{Conclusion}
In this paper, we show how FCA can help clustering of interestingness measures. In general, we can notice that FCA validates the clustering results (for both \textit{AHC and K-means}). In fact, the verification of some clusters, like \textit{C1} or \textit{C4} is clear through the lattice, especially when we find the concept which acts as center of the group. However, some exceptions have been encountered. For instance, we have difficulty to discern the \textit{C2} cluster. In another hand, we note the separation of some clusters into two (or three) groups like \textit{C5}. Our approach can be generalized to other data sets. It is mainly based on the visualization mode, and can be inefficient when dealing with very large clusters. It is not scalable considering the size of a cluster. Thus we are currently working on a formalization of our approach that could allow to run the process in a semi-automatic way.


\end{document}